\documentclass[11pt]{article}
\usepackage{amsfonts}
\usepackage{amssymb,amsmath,epsf,graphicx}
\usepackage{cite, natbib}
\usepackage{time}

\newcommand{\standardthanksIFM}{This work was supported by the William Dow Chair in Political Economy (McGill University), the Canada Research Chair Program (Chair in Econometrics, Universit\'{e} de Montr\'{e}al), the Bank of Canada (Research Fellowship), a Guggenheim  Fellowship, a Konrad-Adenauer  Fellowship (Alexander-von-Humboldt Foundation, Germany), the Institut de finance math\'{e}matique de Montr\'{e}al (IFM2), the Canadian Network of Centres of Excellence [program on \emph{Mathematics of Information Technology and Complex Systems} (MITACS)], the Natural Sciences and Engineering Research Council of Canada, the Social Sciences and Humanities Research Council of Canada, and the Fonds de recherche sur la soci\'{e}t\'{e} et la culture (Qu\'{e}bec).}
\newcommand{\DufourAddress}{William Dow Professor of Economics, McGill University,
  Centre interuniversitaire de recherche en analyse des organisations (CIRANO), and Centre interuniversitaire de recherche en   \'{e}conomie quantitative (CIREQ).
  Mailing address:  Department of Economics, McGill University, Leacock Building, Room 919,  855 Sherbrooke Street West, Montr\'{e}al, Qu\'{e}bec H3A 2T7, Canada.
  TEL: (1) 514 398 4400 ext. 09156; FAX: (1) 514 398 4800; e-mail: jean-marie.dufour@mcgill.ca. Web page:  http://www.jeanmariedufour.com}
\newcommand{\LugerAddress}{D\'{e}partement de finance, assurance et immobilier, Universit\'{e} Laval, Qu\'{e}bec, Qu\'{e}bec G1V 0A6, Canada. E-mail address: richard.luger@fsa.ulaval.ca.}
\providecommand{\vartitleadjust}{}
\renewcommand{\vartitleadjust}{}

\providecommand{\varthanks}{\standardthanksIFM}
\renewcommand{\varthanks}{\standardthanksIFM}
\providecommand{\thanksvar}{\thanks{\varthanks}}
\renewcommand{\thanksvar}{\thinspace \thanks{\ \ \varthanks}}

\providecommand{\vartitle}{\vartitleadjust Identification-robust moment-based tests for Markov-switching \\
in autoregressive models}
\renewcommand{\vartitle}{\vartitleadjust Identification-robust moment-based tests for Markov-switching \\
in autoregressive models}
\providecommand{\varAuthors}{Jean-Marie Dufour \thanks{\ \ \DufourAddress} \\
 McGill University \and 
 Richard Luger \thanks{\ \ \LugerAddress} \\ 
 Universit\'e Laval}
\renewcommand{\varAuthors}{Jean-Marie Dufour \thanks{\ \ \DufourAddress} \\
 McGill University \and 
 Richard Luger \thanks{\ \ \LugerAddress} \\ 
 Universit\'e Laval}
\providecommand{\vardate}{\today, \texttime}
\renewcommand{\vardate}{\today}

\input{tcilatex}

\begin{document}

\title{\vartitle
\thanksvar }
\author{\varAuthors}
\date{\vardate}
\maketitle

\thispagestyle{empty}

\renewcommand{\baselinestretch}{1.55}

\newpage \thispagestyle{empty}

\begin{center}
{\ \textbf{ABSTRACT} }

\quad
\end{center}

\noindent

This paper develops tests of the null hypothesis of linearity in the context
of autoregressive models with Markov-switching means and variances. These
tests are robust to the identification failures that plague conventional
likelihood-based inference methods. The approach exploits the moments of
normal mixtures implied by the regime-switching process and uses Monte Carlo
test techniques to deal with the presence of an autoregressive component in
the model specification. The proposed tests have very respectable power in
comparison to the optimal tests for Markov-switching parameters of %
\citet{Carrasco-Hu-Ploberger:2014} and they are also quite attractive owing
to their computational simplicity. The new tests are illustrated with an
empirical application to an autoregressive model of U.S. output growth.

{\ \bigskip }


{\ \bigskip }


{\ \newpage }

{\ \setcounter{page}{1} }

{\ \renewcommand{\thefootnote}{\arabic{footnote}} \setcounter{footnote}{0} }

\section{Introduction}

The extension of the linear autoregressive model proposed by %
\citet{Hamilton:1989} allows the mean and variance of a time series to
depend on the outcome of a latent process, assumed to follow a Markov chain.
The evolution over time of the latent state variable gives rise to an
autoregressive process with a mean and variance that switch according to the
transition probabilities of the Markov chain. \citet{Hamilton:1989} applies
the Markov-switching model to U.S. output growth rates and argues that it
encompasses the linear specification. This class of models has also been
used to model potential regime shifts in foreign exchange rates %
\citep{Engel-Hamilton:1990}, stock market volatility %
\citep{Hamilton-Susmel:1994}, real interest rates \citep{Garcia-Perron:1996}%
, corporate dividends \citep{Timmermann:2001}, the term structure of
interest rates \citep{Ang-Bekaert:2002a}, portfolio allocation %
\citep{Ang-Bekaert:2002b}, and government policy \citep{Davig:2004}. A
comprehensive treatment of Markov-switching models and many references are
found in \citet{Kim-Nelson:1999}, and more recent surveys of this class of
models are provided by \citet{Guidolin:2011} and \citet{Hamilton:2016}.

A fundamental question in the application of such models is whether the
data-generating process is indeed characterized by regime changes in its
mean or variance. Statistical testing of this hypothesis poses serious
difficulties for conventional likelihood-based methods because two important
assumptions underlying standard asymptotic theory are violated under the
null hypothesis of no regime change. Indeed, if a two-regime model is fitted
to a single-regime linear process, the parameters which describe the second
regime are unidentified. Moreover, the derivative of the likelihood function
with respect to the mean and variance are identically zero when evaluated at
the constrained maximum under both the null and alternative hypotheses.
These difficulties combine features of the statistical problems discussed in %
\citet{Davies:1977, Davies:1987}, \citet{Watson-Engle:1985}, and %
\citet{Lee-Chesher:1986}. The end result is that the information matrix is
singular under the null hypothesis, and the usual likelihood-ratio test does
not have an asymptotic chi-squared distribution in this case. Conventional
likelihood-based inference in the context of Markov-switching models can
thus be very misleading in practice. Indeed, the simulation results reported
by \citet{Psaradakis-Sola:1998} reveal just how poor the first-order
asymptotic approximations to the finite-sample distribution of the
maximum-likelihood estimates can be.

\citet{Hansen:1992, Hansen:1996} and \citet{Garcia:1998} proposed
likelihood-ratio tests specifically tailored to deal with the kind of
violations of the regularity conditions which arise in Markov-switching
models. Their methods differ in terms of which parameters are considered of
interest and those taken as nuisance parameters. Both methods require a
search over the intervening nuisance parameter space with an evaluation of
the Markov-switching likelihood function at each considered grid point,
which makes them computationally expensive. %
\citet{Carrasco-Hu-Ploberger:2014} derive asymptotically optimal tests for
Markov-switching parameters. These information matrix-type tests only
require estimating the model under the null hypothesis, which is a clear
advantage over \citet{Hansen:1992, Hansen:1996} and \citet{Garcia:1998}.
However, the asymptotic distribution of the optimal tests is not free of
nuisance parameters, so \citet{Carrasco-Hu-Ploberger:2014} suggest a
parametric bootstrap procedure to find the critical values.

In this paper, we propose new tests for Markov-switching models which, just
like the \citet{Carrasco-Hu-Ploberger:2014} tests, circumvent the
statistical problems and computational costs of likelihood-based methods.
Specifically, we \emph{first} propose computationally simple test statistics
-- based on least-squares residual moments -- for the hypothesis of no
Markov-switching (or linearity) in autoregressive models. The residual
moment statistics considered include statistics focusing on the mean,
variance, skewness, and excess kurtosis of estimated least-squares
residuals. The different statistics are combined through the minimum or the
product of approximate marginal $p$-values.

\emph{Second}, we exploit the computational simplicity of the test
statistics to obtain exact and asymptotically valid test procedures, which
do not require deriving the asymptotic distribution of the test statistics
and automatically deal with the identification difficulties associated with
such models. Even if the distributions of these combined statistics may be
difficult to establish analytically, the level of the corresponding test is
perfectly controlled. This is made possible through the use of \emph{Monte
Carlo} (MC) test methods. When no new nuisance parameter appears in the null
distribution of the test statistic, such methods allow one to control
perfectly the level of a test, irrespective of the distribution of the test
statistic, as long as the latter can be simulated under the null hypothesis;
see {\citet{Dwass:1957}, \citet{Barnard:1963}, \citet{Birnbaum:1974}, and %
\citet{Dufour:2006}. This feature holds for a fixed number of replications,
which can be quite small. For example, $19$ replications of the test
statistic are sufficient to obtain a test with exact level ${.05}$. A larger
number of replications decreases the sensitivity of the test to the
underlying randomization and typically leads to power gains. %
\citet{Dufour-Khalaf-Bernard-Genest:2004}, however, find that increasing the
number of replications beyond $100$ has only a small effect on power. }

Further, when nuisance parameters are present -- as in the case of linearity
tests studied here -- the procedure can be extended through the use of \emph{%
maximized Monte Carlo}{\ (MMC) tests \citep{Dufour:2006}. Two variants of
this procedure are described: a fully exact version which requires
maximizing a $p$-value function over the nuisance parameter space under the
null hypothesis (here, the autoregressive coefficients), and an approximate
one based on a (potentially much smaller) consistent set estimator of the
autoregressive parameters. Both procedures are valid (in finite samples or
asymptotically) without any need to establish the asymptotic distribution of
the fundamental test statistics (here residual moment-based statistics) or
the convergence of the empirical distribution of the simulated test
statistics toward the asymptotic distribution of the fundamental test
statistic used (as in bootstrapping). }

When the nuisance-parameter set on which the $p$-values are computed is
reduced to a single point -- a consistent estimator of the nuisance
parameters under the null hypothesis -- the MC test can be interpreted as a
parametric bootstrap. The implementation of this type of procedure is also
considerably simplified through the use of our moment-based test statistics.
It is important to emphasize that evaluating the $p$-value function is far
simpler to do than computing the likelihood function of the Markov-switching
model, as required by the methods of \citet{Hansen:1992, Hansen:1996} and %
\citet{Garcia:1998}. The MC tests are also far simpler to compute than the
information matrix-type tests of \citet{Carrasco-Hu-Ploberger:2014}, which
require a grid search for a supremum-type statistic (or numerical
integration for an exponential-type statistic) over \emph{a priori} measures
of the distance between potentially regime-switching parameters and another
parameter characterizing the serial correlation of the Markov chain under
the alternative.

\emph{Third,} we conduct simulation experiments to examine the performance
of the proposed tests using the optimal tests of %
\citet{Carrasco-Hu-Ploberger:2014} as the benchmark for comparisons. The new
moment-based tests are found to perform remarkably well when compared to the
asymptotically optimal ones, especially when the variance is subject to
regime changes. Finally, the proposed methods are illustrated by revisiting
the question of whether U.S. real GNP growth can be described as an
autoregressive model with Markov-switching means and variances using the
original \citet{Hamilton:1989} data set from 1952 to 1984, as well as an
extended data set from 1952 to 2010. We find that the empirical evidence
does not justify a rejection of the linear model over the period 1952--1984.
However, the linear autoregressive model is firmly rejected over the
extended time period.

The paper is organized as follows. Section 2 describes the autoregressive
model with Markov-switching means and variances. Section 3 presents the
moments of normal mixtures implied by the regime-switching process and the
test statistics we propose to combine for capturing those moments. Section 3
also explains how the MC test techniques can be used to deal with the
presence of an autoregressive component in the model specification. Section
4 examines the performance of the developed MC tests in simulation
experiments using the optimal tests for Markov-switching parameters of %
\citet{Carrasco-Hu-Ploberger:2014} as the benchmark for comparison purposes.
Section 5 then presents the results of the empirical application to U.S.
output growth and Section 6 concludes.

\section{Markov-switching model}

We consider an autoregressive model with Markov-switching means and
variances defined by 
\begin{equation}
y_{t}=\mu _{s_{t}}+\sum_{k=1}^{r}\phi _{k}(y_{t-k}-\mu _{s_{t-k}})+\sigma
_{s_{t}}\varepsilon _{t}  \label{mini}
\end{equation}%
where the innovation terms $\{\varepsilon _{t}\}$ are independently and
identically distributed (i.i.d.) according to the $N(0,1)$ distribution. The
time-varying mean and variance parameters of the observed variable $y_{t}$
are functions of a latent first-order Markov chain process $\{S_{t}\}.$ The
unobserved random variable $S_{t}$ takes integer values in the set $\{1,2\}$
such that \sloppy$\Pr (S_{t}=j)=\sum_{i=1}^{2}p_{ij}\Pr (S_{t-1}=i)$, with $%
p_{ij}=\Pr (S_{t}=j\,|\,S_{t-1}=i)$. The one-step transition probabilities
are collected in the matrix 
\begin{equation*}
\mathbf{P}=\left[ 
\begin{array}{cc}
p_{11} & p_{12} \\ 
p_{21} & p_{22}%
\end{array}%
\right] 
\end{equation*}%
where $\sum_{j=1}^{2}p_{ij}=1$\thinspace $,$ for $i=1,2$. Furthermore, $S_{t}
$ and $\varepsilon _{\tau }$ are assumed independent for all $t,\tau $.

The model in (\ref{mini}) can also be conveniently expressed as 
\begin{equation}
y_{t}=\sum_{i=1}^{2}\mu _{i}\mathbb{I}[S_{t}=i]+\sum_{k=1}^{r}\phi _{k}\big(%
y_{t-k}-\sum_{i=1}^{2}\mu _{i}\mathbb{\mathbb{I}}[S_{t-k}=i]\big)%
+\sum_{i=1}^{2}\sigma _{i}\mathbb{\mathbb{I}}[S_{t}=i]\varepsilon _{t}
\label{m1}
\end{equation}%
where $\mathbb{I}[A]$ is the indicator function of event $A$, which is equal
to 1 when $A$ occurs and 0 otherwise. Here $\mu _{i}$ and $\sigma _{i}^{2}$
are the conditional mean and variance given the regime $S_{t}=i.$

The model parameters are collected in the vector $\boldsymbol{\theta }=(\mu
_{1},\mu _{2},\sigma _{1},\sigma _{2},\phi _{1},\ldots ,\,\phi
_{r},p_{11},p_{22})^{\prime }.$ The sample (log) likelihood, conditional on
the first $r$ observations of $y_{t}$, is then given by 
\begin{equation}
L_{T}(\boldsymbol{\theta })=\log f(\boldsymbol{y}_{1}^{T}\,|\,\boldsymbol{y}%
_{-r+1}^{0};\boldsymbol{\theta })=\sum_{t=1}^{T}\log f(y_{t}\,|\,\boldsymbol{%
y}_{-r+1}^{t-1};\boldsymbol{\theta })  \label{like}
\end{equation}%
where $\boldsymbol{y}_{-r+1}^{t}=\{y_{-r+1},\ldots ,\,y_{t}\}$ denotes the
sample of observations up to time $t$, and 
\begin{equation*}
f(y_{t}\,|\,\boldsymbol{y}_{-r+1}^{t-1};\boldsymbol{\theta }%
)=\sum_{s_{t}=1}^{2}\sum_{s_{t-1}=1}^{2}...%
\sum_{s_{t-r}=1}^{2}f(y_{t},S_{t}=s_{t},S_{t-1}=s_{t-1},\ldots
,\,S_{t-r}=s_{t-r}\,|\,\boldsymbol{y}_{-r+1}^{t-1};\boldsymbol{\theta })\,.
\end{equation*}%
\citet{Hamilton:1989} proposes an algorithm for making inferences about the
unobserved state variable $S_{t}$ given observations on $y_{t}.$ His
algorithm also yields an evaluation of the sample likelihood in (\ref{like}%
), which is needed to find the maximum likelihood (ML) estimates of $%
\boldsymbol{\theta }$.

The sample likelihood $L_{T}(\boldsymbol{\theta })$ in (\ref{like}) has
several unusual features which make it notoriously difficult for standard
optimizers to explore. In particular, the likelihood function has several
modes of equal height. These modes correspond to the different ways of
reordering the state labels. There is no difference between the likelihood
for $\mu _{1}=\mu _{1}^{\ast }$ , $\mu _{2}=\mu _{2}^{\ast }$, $\sigma
_{1}=\sigma _{1}^{\ast }$, $\sigma _{2}=\sigma _{2}^{\ast }$ and the
likelihood for $\mu _{1}=\mu _{2}^{\ast }$ , $\mu _{2}=\mu _{1}^{\ast }$, $%
\sigma _{1}=\sigma _{2}^{\ast }$, $\sigma _{2}=\sigma _{1}^{\ast }$. %
\citet[][Ch. 1]{Rossi:2014} provides a nice discussion of these issues in
the context of normal mixtures, which is a special case implied by (\ref{m1}%
) when the $\phi $'s are zero. He shows that the likelihood has numerous
points where the function is not defined with an infinite limit.
Furthermore, the likelihood function also has saddle points containing local
maxima. This means that standard numerical optimizers are likely to converge
to a local maximum and will therefore need to be started from several points
in a constrained parameter space in order to find the ML estimates.

\section{Tests of linearity}

The Markov-switching model in (\ref{m1}) nests the following linear
autoregressive (AR) specification as a special case: 
\begin{equation}
y_{t} = c + \sum_{k=1}^{r}\phi _{k} y_{t-k} +\sigma_1 \varepsilon _{t},
\label{AR}
\end{equation}
where $c=\mu_1(1-\sum_{k=1}^r \phi_k)$. Here $\mu_1$ and $\sigma_1^2$ refer
to the single-regime mean and variance parameters. It is well known that the
conditional ML estimates of the linear model can be obtained from an
ordinary least squares (OLS) regression \citep[][Ch. 5]{Hamilton:1994}. A
problem with the ML approach is that the likelihood function will always
increase when moving from the linear model in (\ref{AR}) to the two-regime
model in (\ref{m1}) as any increase in flexibility is always rewarded. In
order to avoid over-fitting, it is therefore desirable to test whether the
linear specification provides an adequate description of the data.

Given model (\ref{m1}), the null hypothesis of linearity can be expressed as
either $(\mu _{1}=\mu _{2},$ $\sigma _{1}=\sigma _{2})$ or $(p_{11}=1,$ $%
p_{21}=1)$ or $(p_{12}=1,$ $p_{22}=1).$ It is easy to see that if $%
(\mu_{1}=\mu _{2},$ $\sigma _{1}=\sigma _{2})$, then the transition
probabilities are unidentified. On the contrary, if $(p_{11}=1,$ $p_{21}=1)$
then it is $\mu _{2}$ and $\sigma _{2}$ which become unidentified, whereas
if $(p_{12}=1,$ $p_{22}=1)$ then $\mu _{1}$ and $\sigma _{1}$ become
unidentified. One of the regularity conditions underlying the usual
asymptotic distributional theory of ML estimates is that the information
matrix be nonsingular; see, for example, \citet[][Ch.
7]{Gourieroux-Monfort:1995}. Under the null hypothesis of linearity, this
condition is violated since the likelihood function in (\ref{like}) is flat
with respect to the unidentified parameters at the optimum. A singular
information matrix results also from another, less obvious, problem: the
derivatives of the likelihood function with respect to the mean and variance
are identically zero when evaluated at the constrained maximum; see %
\citet{Hansen:1992} and \citet{Garcia:1998}.

\subsection{Mixture model}

We begin by considering the mean-variance switching model: 
\begin{equation}
y_{t}=\mu _{1}\mathbb{I}[S_{t}=1]+\mu _{2}\mathbb{I}[S_{t}=2]+\big(\sigma
_{1}\mathbb{I}[S_{t}=1]+\sigma _{2}\mathbb{I}[S_{t}=2]\big)\varepsilon _{t},
\label{m0}
\end{equation}%
where $\varepsilon _{t}\sim $ i.i.d. $N(0,1)$. The Markov chain governing $%
S_{t}$ is assumed ergodic and we denote the ergodic probability associated
with state $i$ by $\pi _{i}.$ Note that a two-state Markov chain is ergodic
provided that $p_{11}<1,$ $p_{22}<1,$ and $p_{11}+p_{22}>0$ \citep[][p.
683]{Hamilton:1994}. As we already mentioned, the null hypothesis of
linearity (no regime changes) can be expresses as 
\begin{equation*}
H_{0}(\mu ,\sigma ):\mu _{1}=\mu _{2}\text{ and }\sigma _{1}=\sigma _{2},
\end{equation*}%
and a relevant alternative hypothesis states that the mean and/or variance
is subject to first-order Markov-switching. The tests of $H_{0}(\mu ,\sigma )
$ we develop exploit the fact that the marginal distribution of $y_{t}$ is a
mixture of two normal distributions. Indeed, under the maintained assumption
of an ergodic Markov chain we have: 
\begin{equation}
y_{t}\sim \pi _{1}N(\mu _{1},\sigma _{1}^{2})+\pi _{2}N(\mu _{2},\sigma
_{2}^{2}),  \label{mix}
\end{equation}%
where $\pi _{1}=(1-p_{22})/(2-p_{11}-p_{22})$ and $\pi _{2}=1-\pi _{1}$. In
the spirit of \cite{Cho-White:2007} and \citet{Carter-Steigerwald:2012,
Carter-Steigerwald:2013}, the suggested approach ignores the Markov property
of $S_{t}$.

The marginal distribution of $y_{t}$ given in (\ref{mix}) is a weighted
average of two normal distributions. \citet{Timmermann:2000} shows that the
mean ($\mu $), unconditional variance ($\sigma ^{2}$), skewness coefficient (%
$\sqrt{b_{1}}$), and excess kurtosis coefficient ($b_{2}$) associated with (%
\ref{mix}) are given by 
\begin{eqnarray}
\mu &=&\pi _{1}\mu _{1}+\pi _{2}\mu _{2},  \label{mean} \\[3ex]
\sigma ^{2} &=&\pi _{1}\sigma _{1}^{2}+\pi _{2}\sigma _{2}^{2}+\pi _{1}\pi
_{2}(\mu _{2}-\mu _{1})^{2},  \label{var} \\[3ex]
\sqrt{b_{1}} &=&\frac{\pi _{1}\pi _{2}(\mu _{1}-\mu _{2})\big\{%
3(\sigma_{1}^{2}-\sigma _{2}^{2})+(1-2\pi _{1})(\mu _{2}-\mu _{1}^2)^{2}%
\big\}}{\big(\pi _{1}\sigma _{1}^{2}+\pi _{2}\sigma _{2}^{2}+\pi _{1}\pi
_{2}(\mu_{2}-\mu_{1})^{2}\big)^{3/2}},  \label{skewness} \\[3ex]
b_{2} &=&\frac{a}{b},  \label{kurtosis}
\end{eqnarray}
where 
\begin{equation*}
\begin{array}{l}
a=3\pi _{1}\pi _{2}(\sigma _{2}^{2}-\sigma _{1}^{2})^{2}+6(\mu _{2}-\mu
_{1})^{2}\pi _{1}\pi _{2}(2\pi _{1}-1)(\sigma _{2}^{2}-\sigma _{1}^{2}) \\%
[3ex] 
\;\;\;\;\;\;\;+\pi _{1}\pi _{2}(\mu _{2}-\mu _{1})^{4}(1-6\pi _{1}\pi _{2}),
\\[3ex] 
b=\big(\pi _{1}\sigma _{1}^{2}+\pi _{2}\sigma _{2}^{2}+\pi _{1}\pi _{2}(\mu
_{2}-\mu _{1})^{2}\big)^{2}.%
\end{array}%
\end{equation*}

When compared to a bell-shaped normal distribution, the expressions in (\ref%
{mean})--(\ref{kurtosis}) imply that a mixture distribution can be
characterized by any of the following features: the presence of two peaks,
right or left skewness, or excess kurtosis. The extent to which these
characteristics will be manifest depends on the relative values of $\pi _{1}$
and $\pi _{2}$ by which the component distributions in (\ref{mix}) are
weighted, and on the distance between the component distributions. This
distance can be characterized by either the separation between the
respective means, $\Delta _{\mu }=\mu _{2}-\mu _{1}$, or by the separation
between the respective standard deviations, $\Delta _{\sigma }=\sigma _{2}-
\sigma _{1}$, where we adopt the convention that $\mu _{2}>\mu _{1}$ and $%
\sigma _{2}>\sigma _{1}$. For example, if $\Delta _{\sigma }=0$, then the
skewness and relative difference between the two peaks of the mixture
distribution depends on $\Delta _{\mu }$ and the weights $\pi _{1}$ and $\pi
_{2}.$ When $\pi _{1}=\pi _{2}$, the mixture distribution is symmetric with
two modes becoming more distinct as $\Delta _{\mu }$ increases. On the
contrary, if $\Delta _{\mu }=0$ then the mixture distribution will have
heavy tails depending on the difference between the component standard
deviations and their relative weights. See \citet[][Ch. 22]{Hamilton:1994}, %
\citet{Timmermann:2000}, and \citet[][Ch. 1]{Rossi:2014} for more on these
effects.

To test $H_{0}(\mu ,\sigma )$, we propose a combination of four test
statistics based on the theoretical moments in (\ref{mean})--(\ref{kurtosis}%
). The four individual statistics are computed from the residual vector $%
\hat{\boldsymbol{\varepsilon }}=(\hat{\varepsilon}_{1},\hat{\varepsilon}%
_{2},\ldots ,\,\hat{\varepsilon}_{T})^{\prime }$ comprising the residuals $%
\hat{\varepsilon}_{t}=y_{t}-\bar{y}$, themselves computed as the deviations
from the sample mean. Each statistic is meant to detect a specific
characteristic of mixture distributions. The first of these statistics is 
\begin{equation}
M(\hat{\boldsymbol{\varepsilon }})=\frac{|m_{2}-m_{1}|}{\sqrt{%
s_{2}^{2}+s_{1}^{2}}},  \label{m}
\end{equation}%
where 
\begin{equation*}
m_{2}=\frac{\sum_{t=1}^{T}\hat{\varepsilon}_{t}\mathbb{I}[\hat{\varepsilon}%
_{t}>0]}{\sum_{t=1}^{T}\mathbb{I}[\hat{\varepsilon}_{t}>0]},\qquad s_{2}^{2}=%
\frac{\sum_{t=1}^{T}(\hat{\varepsilon}_{t}-m_{2})^{2}\mathbb{I}[\hat{%
\varepsilon}_{t}>0]}{\sum_{t=1}^{T}\mathbb{I}[\hat{\varepsilon}_{t}>0]},
\end{equation*}%
and 
\begin{equation*}
m_{1}=\frac{\sum_{t=1}^{T}\hat{\varepsilon}_{t}\mathbb{I}[\hat{\varepsilon}%
_{t}<0]}{\sum_{t=1}^{T}\mathbb{I}[\hat{\varepsilon}_{t}<0]},\qquad s_{1}^{2}=%
\frac{\sum_{t=1}^{T}(\hat{\varepsilon}_{t}-m_{1})^{2}\mathbb{I}[\hat{%
\varepsilon}_{t}<0]}{\sum_{t=1}^{T}\mathbb{I}[\hat{\varepsilon}_{t}<0]}.
\end{equation*}%
The statistic in (\ref{m}) is a standardized difference between the means of
the observations situated above the sample mean and those below the sample
mean. The next statistic partitions the observations on the basis of the
sample variance $\hat{\sigma}^{2}=T^{-1}\sum_{t=1}^{T}\hat{\varepsilon}%
_{t}^{2}.$ Specifically, we consider 
\begin{equation}
V(\hat{\boldsymbol{\varepsilon }})=\frac{v_{2}(\hat{\boldsymbol{\varepsilon }%
})}{v_{1}(\hat{\boldsymbol{\varepsilon }})},  \label{v}
\end{equation}%
where 
\begin{equation*}
v_{2}=\frac{\sum_{t=1}^{T}\hat{\varepsilon}_{t}^{2}\mathbb{I}[\hat{%
\varepsilon}_{t}^{2}>\hat{\sigma}^{2}]}{\sum_{t=1}^{T}\mathbb{I}[\hat{%
\varepsilon}_{t}^{2}>\hat{\sigma}^{2}]},\qquad v_{1}=\frac{\sum_{t=1}^{T}%
\hat{\varepsilon}_{t}^{2}\mathbb{I}[\hat{\varepsilon}_{t}^{2}<\hat{\sigma}%
^{2}]}{\sum_{t=1}^{T}\mathbb{I}[\hat{\varepsilon}_{t}^{2}<\hat{\sigma}^{2}]},
\end{equation*}%
so that $v_{2}>v_{1}.$ Note that we partition on the basis of average values
because (\ref{mix}) is a two-component mixture. The last two statistics are
the absolute values of the coefficients of skewness and excess kurtosis: 
\begin{equation}
S(\hat{\boldsymbol{\varepsilon }})=\left\vert \;\frac{\sum_{t=1}^{T}\hat{%
\varepsilon}_{t}^{3}}{T(\hat{\sigma}^{2})^{3/2}}\;\right\vert   \label{sk}
\end{equation}%
and 
\begin{equation}
K(\hat{\boldsymbol{\varepsilon }})=\left\vert \;\frac{\sum_{t=1}^{T}\hat{%
\varepsilon}_{t}^{4}}{T(\hat{\sigma}^{2})^{2}}-3\;\right\vert ,  \label{ku}
\end{equation}%
which were also considered in \cite{Cho-White:2007}. Observe that the
statistics in (\ref{m})--(\ref{ku}) can only be non-negative and are each
likely to be larger in value under the alternative hypothesis. Taken
together, they constitute a potentially useful battery of statistics to test 
$H_{0}(\mu ,\sigma )$ by capturing characteristics of the first four moments
of normal mixtures. As one would expect, the power of the tests based on (%
\ref{m})--(\ref{ku}) will generally be increasing with the frequency of
regime changes.

It is easy to see that the statistics in (\ref{m})--(\ref{ku}) are exactly
pivotal as they all involve ratios and can each be computed from the vector
of standardized residuals $\hat{\boldsymbol{\varepsilon }}/\hat{\sigma}$,
which are scale and location invariant under the null of linearity. That is,
the vector of statistics $(M(\hat{\boldsymbol{\varepsilon }}),V(\hat{%
\boldsymbol{\varepsilon }}),S(\hat{\boldsymbol{\varepsilon }}),K(\hat{%
\boldsymbol{\varepsilon }}))^{\prime }$ is distributed like \sloppy$\big(M(%
\hat{\boldsymbol{\eta }}),V(\hat{\boldsymbol{\eta }}),S(\hat{\boldsymbol{%
\eta }}),K(\hat{\boldsymbol{\eta }})\big)^{\prime }$, where $\boldsymbol{%
\eta }\sim N(0,I_{T})$ and $\hat{\boldsymbol{\eta }}=\boldsymbol{\eta }-\bar{%
\boldsymbol{\eta }}$. The null distribution of the proposed test statistics
can thus be simulated to any degree of precision, thereby paving the way for
an MC test as follows. 

First, compute each of the statistics in (\ref{m})--(\ref{ku}) with the
actual data to obtain $(M(\hat{\boldsymbol{\varepsilon }}),V(\hat{%
\boldsymbol{\varepsilon }}),S(\hat{\boldsymbol{\varepsilon }}),K(\hat{%
\boldsymbol{\varepsilon }}))^{\prime }.$ Then generate $N-1$ mutually
independent $T\times 1$ vectors $\boldsymbol{\eta }_{i},$ $i=1,\ldots ,\,N-1,
$ where $\boldsymbol{\eta }_{i}\sim N(0,I_{T}).$ For each such vector
compute $\hat{\boldsymbol{\eta }}_{i}=(\hat{\eta}_{i1},\hat{\eta}%
_{i2},\ldots ,\,\hat{\eta}_{iT})^{\prime }$ with typical element $\hat{\eta}%
_{it}=\eta _{it}-\overline{\eta }_{i},$ where $\overline{\eta }_{i}$ is the
sample mean, and compute the statistics in (\ref{m})--(\ref{ku}) based on $%
\hat{\boldsymbol{\eta }}_{i}$ so as to obtain $N-1$ statistics vectors $(M(%
\hat{\boldsymbol{\eta }}_{i}),V(\hat{\boldsymbol{\eta }}_{i}),S(\hat{%
\boldsymbol{\eta }}_{i}),K(\hat{\boldsymbol{\eta }}_{i}))^{\prime },$ $%
i=1,\ldots ,\,N-1.$ Let $\xi $ denote any one of the above four statistics, $%
\xi _{0}$ its original data-based value, and $\xi _{i},$ $i=1,\ldots ,\,N-1,$
the corresponding simulated values. The individual MC $p$-values are then
given by 
\begin{equation}
G_{\xi }[\xi _{0};N]=\frac{N+1-R_{\xi }[\xi _{0};N]}{N},  \label{p-value}
\end{equation}%
where $R_{\xi }[\xi _{0};N]$ is the rank of $\xi _{0}$ when $\xi _{0},\xi
_{1},\ldots ,\,\xi _{N-1}$ are placed in increasing order. The associated MC
critical regions are defined as 
\begin{equation*}
W_{N}^{(\xi )}=\big\{R_{\xi }[\xi _{0};N]\geq c_{N}(\alpha _{\xi })\big\}
\end{equation*}%
with 
\begin{equation*}
c_{N}(\alpha _{\xi })=N-I[N\alpha _{\xi }]+1,
\end{equation*}%
where $I[x]$ denotes the largest integer not exceeding $x$. These MC
critical regions are exact for any given sample size, $T$. Further
discussion and applications of the MC test technique can be found in %
\citet{Dufour-Khalaf:2001} and \citet{Dufour:2006}.

Note that the MC $p$-values $G_{M}[M(\hat{\boldsymbol{\varepsilon }});N],$ $%
G_{V}[V(\hat{\boldsymbol{\varepsilon }});N],$ $G_{S}[S(\hat{\boldsymbol{%
\varepsilon }});N],$ and $G_{K}[K(\hat{\boldsymbol{\varepsilon }});N]$ are
not statistically independent and may in fact have a complex dependence
structure. Nevertheless, if we choose the individual levels such that $%
\alpha _{M}+\alpha _{V}+\alpha _{S}+\alpha _{K}=\alpha $ then, for $%
TS=\{M,V,S,K\},$ we have by the Boole-Bonferroni inequality: 
\begin{equation*}
\Pr \left( \bigcup_{\xi \in TS}W_{N}^{(\xi )}\right) \leq \alpha ,
\end{equation*}%
so the \textit{induced} test, which consists in rejecting $H_{0}(\mu ,\sigma
)$ when any of the individual tests rejects, has level $\alpha $. For
example, if we set each individual test level at 2.5\%, so that we reject if 
$G_{\xi }[\xi _{0};N]\leq 2.5\%$ for any $\xi \in \{M,V,S,K\},$ then the
overall probability of committing a Type I error does not exceed 10\%. Such
Bonferroni-type adjustments, however, can be quite conservative and lead to
power losses; see \citet{Savin:1984} for a survey of these issues.

In order to resolve these multiple comparison issues, we propose an MC test
procedure based on combining individual $p$-values. The idea is to treat the
combination like any other (pivotal) test statistic for the purpose of MC
resampling. As with double bootstrap schemes \citep[][]{MacKinnon:2009},
this approach can be computationally expensive since it requires a second
layer of simulations to obtain the $p$-value of the combined (first-level) $p
$-values. Here though we can ease the computational burden by using
approximate $p$-values in the first level. A remarkable feature of the MC
test combination procedure is that it remains exact even if the first-level $%
p$-values are only approximate. Indeed, the MC procedure implicitly accounts
for the fact that the $p$-value functions may not be individually exact and
yields an overall $p$-value for the combined statistics which itself is
exact. For this procedure, we make use of approximate distribution functions
taking the simple logistic form: 
\begin{equation}
\hat{F}[x]=\frac{\exp (\hat{\gamma}_{0}+\hat{\gamma}_{1}x)}{1+\exp (\hat{%
\gamma}_{0}+\hat{\gamma}_{1}x)},  \label{logi}
\end{equation}%
whose estimated coefficients are given in Table 1 for selected sample sizes.
These coefficients were obtained by the method of non-linear least squares
(NLS) applied to simulated distribution functions comprising a million draws
for each sample size. The approximate $p$-value of, say, $M(\hat{\boldsymbol{%
\varepsilon }})$ is then computed as $\hat{G}_{M}[M(\hat{\boldsymbol{%
\varepsilon }})]=1-\hat{F}_{M}[M(\hat{\boldsymbol{\varepsilon }})]$, where $%
\hat{F}_{M}[x]$ is given by (\ref{logi}) with associated $\hat{\gamma}$'s
from Table 1. The other $p$-values $\hat{G}_{V},\hat{G}_{S},\hat{G}_{K}$ are
computed in a similar way.

We consider two methods for combining the individual $p$-values. The first
one rejects the null when at least one of the $p$-values is sufficiently
small so that the decision rule is effectively based on the statistic 
\begin{equation}
F_{\min }(\hat{\boldsymbol{\varepsilon }})=1-\min \left\{ \hat{G}_{M}[M(\hat{%
\boldsymbol{\varepsilon }})],\hat{G}_{V}[V(\hat{\boldsymbol{\varepsilon }})],%
\hat{G}_{S}[S(\hat{\boldsymbol{\varepsilon }})],\hat{G}_{K}[K(\hat{%
\boldsymbol{\varepsilon }})]\right\} .  \label{cm}
\end{equation}%
The criterion in (\ref{cm}) was suggested by \citet{Tippett:1931} and %
\citet{Wilkinson:1951} for combining inferences obtained from independent
studies. The second method, suggested by \citet{Fisher:1932} and %
\citet{Pearson:1933}, again for independent test statistics, is based on the
product (rather than the minimum) of the $p$-values: 
\begin{equation}
F_{\times }(\hat{\boldsymbol{\varepsilon }})=1-\hat{G}_{M}[M(\hat{%
\boldsymbol{\varepsilon }})]\times \hat{G}_{V}[V(\hat{\boldsymbol{%
\varepsilon }})]\times \hat{G}_{S}[S(\hat{\boldsymbol{\varepsilon }})]\times 
\hat{G}_{K}[K(\hat{\boldsymbol{\varepsilon }})].  \label{cm2}
\end{equation}%
The MC $p$-value of the combined statistic in (\ref{cm}), for example, is
then given by 
\begin{equation}
G_{F_{\min }}[F_{\min }(\hat{\boldsymbol{\varepsilon }});N]=\frac{%
N+1-R_{F_{\min }}[F_{\min }(\hat{\boldsymbol{\varepsilon }});N]}{N},
\label{pv}
\end{equation}%
where $R_{F_{\min }}[F_{\min }(\hat{\boldsymbol{\varepsilon }});N]$ is the
rank of $F_{\min }(\hat{\boldsymbol{\varepsilon }})$ when $F_{\min }(\hat{%
\boldsymbol{\varepsilon }}),F_{\min }(\hat{\boldsymbol{\eta }}_{1}),\ldots
,\,F_{\min }(\hat{\boldsymbol{\eta }}_{N-1})$ are placed in ascending order.
Although the statistics which enter into the computation of (\ref{cm}) and (%
\ref{cm2}) may have a rather complex dependence structure, the MC $p$-values
computed as in (\ref{pv}) are provably exact. See %
\citet{Dufour-Khalaf-Bernard-Genest:2004} and \citet{Dufour-Khalaf-Voia:2014}
for further discussion and applications of these test combination methods.

\subsection{Autoregressive dynamics}

In this section we extend the proposed MC tests to Markov-switching models
with state-independent autoregressive dynamics. To keep the presentation
simple, we describe in detail the test procedure in the case of models with
a first-order autoregressive component. Models with higher-order
autoregressive components are dealt with by a straightforward extension of
the AR(1) case. For convenience, the Markov-switching model with AR(1)
component that we treat is given here as 
\begin{equation}
y_{t}=\mu _{s_{t}}+\phi (y_{t-1}-\mu _{s_{t-1}})+\sigma _{s_{t}}\varepsilon
_{t}  \label{ar2}
\end{equation}%
where 
\begin{eqnarray*}
\mu _{s_{t}} &=&\mu _{1}\mathbb{I}[S_{t}=1]+\mu _{2}\mathbb{I}[S_{t}=2], \\
\sigma _{s_{t}} &=&\sigma _{1}\mathbb{I}[S_{t}=1]+\sigma _{2}\mathbb{I}%
[S_{t}=2].
\end{eqnarray*}%
The tests exploit the fact that, given the true value of $\phi $, the
simulation-based procedures of the previous section can be validly applied
to a transformed model. The idea is that if $\phi $ in (\ref{ar2}) were
known we could test whether $z_{t}(\phi )=y_{t}-\phi y_{t-1}$, defined for $%
t=2,\ldots ,\,T$, follows a mixture of at least two normals.

Indeed, when $\mu _{1}\neq \mu _{2}$ ($\mu _{1},\mu _{2}\neq 0$), the random
variable $z_{t}(\phi )$ follows a mixture of two normals (when $\phi =0$),
three normals (when $|\phi |=1$), or four normals otherwise. That is, when $%
\phi y_{t-1}$ is subtracted on both sides of (\ref{ar2}), the result is a
model with a mean that switches between four states according to 
\begin{equation*}
z_{t}(\phi )=\mu _{1}^{\ast }\mathbb{I}[S_{t}^{\ast }=1]+\mu _{2}^{\ast }%
\mathbb{I}[S_{t}^{\ast }=2]+\mu _{3}^{\ast }\mathbb{I}[S_{t}^{\ast }=3]+\mu
_{4}^{\ast }\mathbb{I}[S_{t}^{\ast }=4]+\big(\sigma _{1}\mathbb{I}%
[S_{t}=1]+\sigma _{2}\mathbb{I}[S_{t}=2]\big)\varepsilon _{t}
\end{equation*}%
where 
\begin{equation}
\mu _{1}^{\ast }=\mu _{1}(1-\phi ),\;\mu _{2}^{\ast }=\mu _{2}-\phi \mu
_{1},\;\mu _{3}^{\ast }=\mu _{1}-\phi \mu _{2},\;\mu _{4}^{\ast }=\mu
_{2}(1-\phi )  \label{mustar}
\end{equation}%
and $S_{t}^{\ast }$ is a first-order, four-state Markov chain with
transition probability matrix 
\begin{equation*}
\mathbf{P}=\left[ 
\begin{array}{cccc}
p_{11} & p_{12} & 0 & 0 \\ 
0 & 0 & p_{21} & p_{22} \\ 
p_{11} & p_{12} & 0 & 0 \\ 
0 & 0 & p_{21} & p_{22}%
\end{array}%
\right] .
\end{equation*}%
If $\mu _{1}\neq \mu _{2}$, the quantities in (\ref{mustar}) admit either
two distinct values (when $\phi =0$), three distinct values (when $\phi =1$
or $-1$), or four distinct values otherwise. Under $H_{0}(\mu ,\sigma )$,
the filtered observations $z_{t}(\phi )$, $t=2,\ldots ,\,T$, are i.i.d. when
evaluated at the true value of the autoregressive parameter.

To deal with the fact that $\phi $ in unknown, we use the extension of the
MC test technique proposed in \citet{Dufour:2006} to deal with the presence
of nuisance parameters. Treating $\phi $ as a nuisance parameter means that
the proposed test statistics become functions of $\hat{\varepsilon}_{t}(\phi
)$, where $\hat{\varepsilon}_{t}(\phi )=z_{t}(\phi )-\bar{z}(\phi )$. Let $%
\Omega _{\phi }$ denote the set of admissible values for $\phi $ which are
compatible with the null hypothesis. Depending on the context, the set $%
\Omega _{\phi }$ may be $\mathbb{R}$ itself, the open interval $(-1,1)$, the
closed interval $[-1,1]$, or any other appropriate subset of $\mathbb{R}$.
In light of a minimax argument \citep{Savin:1984}, the null hypothesis may
then be viewed as a union of point null hypotheses, where each point
hypothesis specifies an admissible value for $\phi $. In this case, the
statistic in (\ref{pv}) yields a test of $H_{0}(\mu ,\sigma )$ with level $%
\alpha $ if and only if 
\begin{equation*}
G_{F_{\min }}[F_{\min }(\hat{\boldsymbol{\varepsilon }});N]\leq \alpha
,\;\;\;\;\forall \phi \in \Omega _{\phi }\,,
\end{equation*}%
or, equivalently, 
\begin{equation*}
\sup_{\phi \in \Omega _{\phi }}G_{F_{\min }}[F_{\min }(\hat{\boldsymbol{%
\varepsilon }});N]\leq \alpha \,.
\end{equation*}%
In words, the null is rejected whenever for all admissible values of $\phi $
under the null, the corresponding point null hypothesis is rejected.
Therefore, if $N\alpha $ is an integer, we have under $H_{0}(\mu ,\sigma )$, 
\begin{equation*}
\Pr \left[ \sup \big\{G_{F_{\min }}[F_{\min }(\hat{\boldsymbol{\varepsilon }}%
);N]\;:\;\phi \in \Omega _{\phi }\big\}\leq \alpha \right] \leq \alpha \,,
\end{equation*}%
\emph{i.e.} the critical region $\sup \{G_{F_{\min }}[F_{\min }(\hat{%
\boldsymbol{\varepsilon }});N]\;:\;\phi \in \Omega _{\phi }\}\leq \alpha $
has level $\alpha $. This procedure is called a maximized MC (MMC) test. It
should be noted that the optimization is done over $\Omega _{\phi }$ holding
fixed the values of the simulated $T\times 1$ vectors $\boldsymbol{\eta }%
_{i},$ $i=1,\ldots ,\,N-1,$ with $\boldsymbol{\eta }_{i}\sim N(0,I_{T})$ --
from which the simulated statistics are obtained.

The maximization involved in the MMC test can be numerically challenging for
Newton-type methods since the simulated $p$-value function is discontinuous.
Search methods for non-smooth objectives which do not rely on gradients are
therefore necessary. A computationally simplified procedure can be based on
a consistent set estimator $C_{T}$ of $\phi $; \emph{i.e.}, one for which $%
\lim_{T\rightarrow \infty }\Pr [\phi \in C_{T}]=1$. For example, if $\hat{%
\phi}_{T}$ is a consistent point estimate of $\phi $ and $c$ is any positive
number, then the set 
\begin{equation*}
C_{T}=\big\{\phi \in \Omega _{\phi }\;:\;\Vert \hat{\phi}_{T}-\phi \Vert <c%
\big\}
\end{equation*}%
is a consistent set estimator of $\phi $; \emph{i.e.}, $\lim_{T\rightarrow
\infty }\Pr [\Vert \hat{\phi}_{T}-\phi \Vert <c]=1,$ $\forall c>0$. Under $%
H_{0}(\mu ,\sigma )$, the critical region based on (\ref{pv}) satisfies 
\begin{equation*}
\lim_{T\rightarrow \infty }\Pr \left( \sup \big\{G_{F_{\min }}[F_{\min }(%
\hat{\boldsymbol{\varepsilon }});N]\;:\;\phi \in C_{T}\big\}\leq \alpha
\right) \leq \alpha \,.
\end{equation*}%
The procedure may even be based on the singleton set $C_{T}=\{\hat{\phi}%
_{T}\},$ which yields a local MC (LMC) test based on a consistent point
estimate. See \citet{Dufour:2006} for additional details.

\section{Simulation evidence}

This section presents simulation evidence on the performance of the proposed
MC tests using model (\ref{ar2}) as the data-generating process (DGP). As a
benchmark for comparison purposes, we take the optimal tests for
Markov-switching parameters developed by \citet{Carrasco-Hu-Ploberger:2014}
(CHP). To describe these tests, let $\ell _{t}=\ell _{t}(\boldsymbol{\theta }%
_{0})$ denote the log of the predictive density of the $t$th observation
under the null hypothesis of a linear model. For model (\ref{ar2}), the
parameter vector under the null hypothesis becomes $\boldsymbol{\theta }%
_{0}=(c,\phi ,\sigma ^{2})^{\prime }$ and we have 
\begin{equation*}
\ell _{t}=-\frac{1}{2}\log (2\pi \sigma ^{2})-\frac{(y_{t}-c-\phi
y_{t-1})^{2}}{2\sigma ^{2}}\,.
\end{equation*}%
Let $\hat{\boldsymbol{\theta }}_{0}$ denote the conditional maximum
likelihood estimates under the null hypothesis (which can be obtained by
OLS) and define 
\begin{equation*}
\ell _{t}^{(1)}=\frac{\partial \ell _{t}}{\partial \boldsymbol{\theta }}%
\Big\lvert_{\boldsymbol{\theta }=\hat{\boldsymbol{\theta }}_{0}}\mbox{ and }%
\ell _{t}^{(2)}=\frac{\partial ^{2}\ell _{t}}{\partial \boldsymbol{\theta }%
\partial \boldsymbol{\theta }^{\prime }}\Big\lvert_{\boldsymbol{\theta }=%
\hat{\boldsymbol{\theta }}_{0}}\,.
\end{equation*}%
The CHP information matrix-type tests are calculated with 
\begin{equation*}
\Gamma _{T}^{\ast }=\Gamma _{T}^{\ast }(\boldsymbol{h},\rho )=\sum_{t}\mu
_{2,t}^{\ast }(\boldsymbol{h},\rho )/\sqrt{T}
\end{equation*}%
where 
\begin{equation*}
\mu _{2,t}^{\ast }(\boldsymbol{h},\rho )=\frac{1}{2}\boldsymbol{h}^{\prime }%
\left[ \ell _{t}^{(2)}+\ell _{t}^{(1)}\ell _{t}^{(1)\prime }+2\sum_{s<t}\rho
^{t-s}\ell _{t}^{(1)}\ell _{s}^{(1)\prime }\right] \boldsymbol{h}\,.
\end{equation*}%
Here the elements of vector $\boldsymbol{h}$ are \emph{a priori} measures of
the distance between the corresponding switching parameters under the
alternative hypothesis and the scalar $\rho $ characterizes the serial
correlation of the Markov chain. To ensure identification, the vector $%
\boldsymbol{h}$ needs to be normalized such that $\Vert \boldsymbol{h}\Vert
=1.$ For given values of $\boldsymbol{h}$ and $\rho $, let $\hat{\boldsymbol{%
\varepsilon }}^{\ast }=\hat{\boldsymbol{\varepsilon }}^{\ast }(\boldsymbol{h}%
,\rho )$ denote the residuals of an OLS regression of $\mu _{2,t}^{\ast }(%
\boldsymbol{h},\rho )$ on $\ell _{t}^{(1)}$.

Following the suggestion in CHP, $\boldsymbol{h}$ in the case of model (\ref%
{ar2}) is a 3-vector whose first and third elements (corresponding to a
switching mean and variance) are generated uniformly over the unit sphere,
and $\rho $ takes values in the interval $[\underline{\rho },\bar{\rho}%
]=[-0.7,0.7].$ The nuisance parameters in $\boldsymbol{h}$ and $\rho $ can
be dealt with in two ways. The first is with a supremum-type test statistic: 
\begin{equation*}
\mbox{supTS}=\sup_{\{\boldsymbol{h},\rho \,:\,\Vert \boldsymbol{h}\Vert =1,%
\underline{\rho }<\rho <\bar{\rho}\}}\frac{1}{2}\left( \max \left( 0,\frac{%
\Gamma _{T}^{\ast }}{\sqrt{\hat{\boldsymbol{\varepsilon }}^{\ast \prime }%
\hat{\boldsymbol{\varepsilon }}^{\ast }}}\right) \right) ^{2}
\end{equation*}%
and the second is with an exponential-type statistic (based on an
exponential prior): 
\begin{equation*}
\mbox{expTS}=\int_{\{\Vert \boldsymbol{h}\Vert =1,\underline{\rho }<\rho <%
\bar{\rho}\}}\Psi (\boldsymbol{h},\rho )\,d\boldsymbol{h}\,d\rho 
\end{equation*}%
where 
\begin{equation*}
\Psi (\boldsymbol{h},\rho )=\left\{ 
\begin{array}{ll}
\sqrt{2\pi }\exp \left[ \frac{1}{2}\left( \frac{\Gamma _{T}^{\ast }}{\sqrt{%
\hat{\boldsymbol{\varepsilon }}^{\ast \prime }\hat{\boldsymbol{\varepsilon }}%
^{\ast }}}-1\right) ^{2}\right] \Phi \left( \frac{\Gamma _{T}^{\ast }}{\sqrt{%
\hat{\boldsymbol{\varepsilon }}^{\ast \prime }\hat{\boldsymbol{\varepsilon }}%
^{\ast }}}-1\right)  & \mbox{ if }\hat{\boldsymbol{\varepsilon }}^{\ast
\prime }\hat{\boldsymbol{\varepsilon }}^{\ast }\neq 0\,, \\[2ex]
1 & \mbox{ otherwise.}%
\end{array}%
\right. 
\end{equation*}%
Here $\Phi (\cdot )$ stands for the standard normal cumulative distribution.
CHP suggest using a parametric bootstrap to assess the statistical
significance of these statistics because their asymptotic distributions are
not free of nuisance parameters. This is done by generating data from the
linear AR model with $\hat{\boldsymbol{\theta }}_{0}$ and calculating $%
\mbox{supTS}$ and $\mbox{expTS}$ with each artificial sample. We implemented
this procedure using 500 bootstrap replications.

In the following tables, LMC and MMC stand for the local and maximized MC
procedures, respectively. The first-level $p$-values are computed from the
estimated distribution functions in Table 1, and the subscript
\textquotedblleft $\min $" is used to indicate that the first-level $p$%
-values are combined via their minimum, while the subscript
\textquotedblleft $\times $" indicates that they are combined via their
product. The MC tests were implemented with $N=100$ and the MMC test was
performed by maximizing the MC $p$-value by grid search over an interval
defined by taking two standard errors on each side of $\hat{\phi}_{0}$, the
OLS estimate of $\phi $. The simulation experiments are based on 1000
replications of each DGP configuration.

For a nominal $5\%$ level, Table 2 reports the empirical size (in
percentage) of the LMC, MMC, supTS, and expTS tests for $\phi=0.1$, $0.9$
and $T=100$, $200.$ The MMC tests are seen to perform according to the
developed theory with empirical rejection rates $\le 5\%$ under the null
hypothesis. The LMC tests based on $\hat \phi_0$ perform remarkably well,
revealing an empirical size close to the nominal $5\%$ level in each case.
The same can be said about the bootstrap $\mbox{supTS}$ and $\mbox{expTS}$
tests even though they seem to be less stable than the LMC tests.

Tables 3 and 4 report the empirical power (in percentage) of the tests for $%
\phi =0.1$ and $\phi =0.9$, respectively. The DGP configurations vary the
separation between the means $\Delta \mu =\mu _{2}-\mu _{1}$ and standard
deviations $\Delta \sigma =\sigma _{2}-\sigma _{1}$ as $(\Delta \mu ,\Delta
\sigma )=(2,0)$, $(0,1)$, $(2,2)$; the sample size as $T=100,$ $200$; and
the transition probabilities as $(p_{11},p_{22})=(0.9,0.9),$ $(0.9,0.5),$ $%
(0.9,0.1)$.

As expected, the power of the proposed tests increases with $\Delta _{\mu }$
and $\Delta _{\sigma }$, and the sample size. For given values of $\Delta
_{\mu }$ and $\Delta _{\sigma }$, test power tends to increase with the
frequency of regime switches. For example, when $\Delta _{\mu }=2$ and $%
\Delta _{\sigma }=1$, the power of the MC tests increases when $p_{22}$
decreases (increase) from 0.9 (0.1) to 0.5. Comparing the LMC$_{\min }$ and
MMC$_{\min }$ to LMC$_{\times }$ and MMC$_{\times }$, respectively, reveals
that there is a power gain in most cases from using the product rule to
combine the first-level $p$-values in the MC procedure. Not surprisingly,
the LMC procedures (based on the point estimate $\hat{\phi}_{0}$) have
better power than the MMC procedures, which maximize the MC $p$-value over a
range of admissible values for $\phi $ in order to hedge the risk of
committing a Type I error.

The $\mbox{supTS}$ and $\mbox{expTS}$ generally tend to be more powerful
than the MC tests, particularly when there are regimes only in the mean (%
\emph{e.g.} $\Delta _{\mu }=2$, $\Delta _{\sigma}=0 $). Nevertheless, it is
quite remarkable that the LMC tests have power approaching that of the $%
\mbox{supTS}$ and $\mbox{expTS}$ tests as soon as the variance is also
subject to regime changes. In some cases, the LMC tests even appears to
outperform the optimal CHP tests. For instance this can be observed in the
middle portion of Table 3, where $\Delta _{\mu }=0$, $\Delta _{\sigma}=1 $.
Another important remark is that the proposed moment-based MC tests are far
easier to compute than the information matrix-type bootstrap tests.

\section{Empirical illustration}

In this section, we present an application of our test procedures to the
study by \citet{Hamilton:1989} who suggested modelling U.S. output growth
with a Markov-switching specification as in (\ref{m1}) with $r=4$ and where
only the mean is subject to regime changes. With this model specification,
business cycle expansions and contractions can be interpreted as a process
of switching between states of high and low growth rates. Hamilton estimated
his model by the method of maximum likelihood with quarterly data ranging
from 1952Q2 to 1984Q4. Probabilistic inferences on the state of the economy
were then calculated and compared to the business-cycle dates as established
by the National Bureau of Economic Research. On the basis of simulated
residual autocorrelations, Hamilton argued that his Markov-switching model
encompasses the linear AR(4) specification.

We applied our proposed MC procedures to formally test the linear AR(4)
specification. In this context, the LMC and MMC procedures are based on the
filtered observations 
\begin{equation*}
z_{t}(\boldsymbol{\phi })=y_{t}-\phi _{1}y_{t-1}-\phi _{2}y_{t-2}-\phi
_{3}y_{t-3}-\phi _{4}y_{t-4},
\end{equation*}%
where $y_{t}$ is 100 times the change in the logarithm of U.S. real GNP.
Following \citet{Carrasco-Hu-Ploberger:2014}, we considered Hamilton's
original data set (135 observations of $y_{t}$) and an extended data set
including observations from 1952Q2 to 2010Q4 (239 observations of $y_{t}$).
The $\boldsymbol{\phi }$ values used in $z_{t}(\boldsymbol{\phi })$ for the
LMC procedure are obtained by an OLS regression of $y_{t}$ on a constant and
four of its lags. The MMC test procedure maximizes the MC $p$-value by grid
search over a four-dimensional box defined by taking 2 standard errors on
each side of the OLS parameter estimates. To ensure stationarity of the
solutions, we only considered grid points for which the roots of the
autoregressive polynomial $1-\phi _{1}z-\phi _{2}z^{2}-\phi _{3}z^{3}-\phi
_{4}z^{4}=0$ lie outside the unit circle. The number of MC replications was
set as $N=100$.

Table 5 shows the test results for the LMC and MMC procedures based on the
minimum and product combination rules. For the MMC statistics the table
reports the maximal MC $p$-value, the $\boldsymbol{\phi }$ values that
maximized the $p$-value function, and the smallest modulus of the roots of $%
1-\phi _{1}z-\phi _{2}z^{2}-\phi _{3}z^{3}-\phi _{4}z^{4}=0.$ These points
on the grid with the highest MMC $p$-values can be interpreted as
Hodges-Lehmann-stye estimates of the autoregressive parameters %
\citep{Hodges-Lehmann:1963}. In the case of the LMC statistics, the reported 
$\boldsymbol{\phi }$ values are simply the OLS point estimates.

For Hamilton's data, the results clearly show that the null hypothesis of
linearity cannot be rejected at usual levels of significance. Furthermore,
the retained values of the autoregressive component yield
covariance-stationary representations of output growth. This shows that the
GNP data from 1952 to 1984 is entirely compatible with a linear and
stationary autoregressive model. It is interesting to note from Table 5 that
the MMC$_{\min }$ and MMC$_{\times }$ procedures find $\boldsymbol{\phi }$
values yielding $p$-values = 1 for the period 1952Q2--1984Q4. Our MC tests,
however, reject the stationary linear AR(4) model with $p$-values $\leq 0.06$
over the extended sample period from 1952 to 2010, which agrees with the
findings of \citet{Carrasco-Hu-Ploberger:2014}. The results presented here
are also consistent with the evidence in \citet{Kim-Nelson:1999} and %
\citet{McConnell-Perez-Quiros:2000} about a structural decline in the
volatility of business cycle fluctuations starting in the mid-1980's -- the
so-called \emph{Great Moderation}.

\section{Conclusion}

We have shown how the MC test technique can be used to obtain provably exact
and useful tests of linearity in the context of autoregressive models with
Markov-switching means and variances. The developed procedure is robust to
the identification issues that plague conventional likelihood-based
inference methods, since all the required computations are done under the
null hypothesis. Another advantage of our MC test procedure is that it is
easy to implement and computationally inexpensive.

The suggested test statistics exploit the fact that, under the
Markov-switching alternative, the observations unconditionally follow a
mixture of at least two normal distributions once the autoregressive
component is properly filtered out. Four statistics, each ones meant to
detect a specific feature of normal mixtures, are combined together either
through the minimum or the product of their individual $p$-values. Of
course, one may combine any subset of the proposed test statistics, or even
include others not considered here. As long as the individual statistics are
pivotal under the null of linearity, the proposed MC test procedure will
control the overall size of the combined test.

The provably exact MMC tests require the maximization of a $p$-value
function over the space of admissible values for the autoregressive
parameters. A simplified version (LMC test) limits the maximization to a
consistent set estimator. Strictly speaking, the LMC tests are no longer
exact in finite samples. Nevertheless, the level constraint will be
satisfied asymptotically under much weaker conditions than those typically
required for the bootstrap. In terms of both size and power, the LMC tests
based on a consistent point estimate of the autoregressive parameters were
found to perform remarkably well in comparison to the bootstrap tests of %
\citet{Carrasco-Hu-Ploberger:2014}.

The developed approach can also be extended to allow for non-normal
mixtures. Indeed, it is easy to see that the standardized residuals $\hat{%
\boldsymbol{\varepsilon }}/\hat{\sigma}$ remain pivotal under the null of
linearity as long as $\varepsilon_t$ in (\ref{m0}) has a completely
specified distribution. As in \citet{Beaulieu-Dufour-Khalaf:2007}, the MMC
test technique can be used to further allow the distribution of $%
\varepsilon_t$ to depend on unknown nuisance parameters. Such extensions go
beyond the scope of the present paper and are left for future work.

{\newpage }

\bibliographystyle{chicago}
\bibliography{Dufour_Luger_2016_Refs}

\begin{thebibliography}{}

\bibitem[\protect\citeauthoryear{Ang and Bekaert}{Ang and
  Bekaert}{2002a}]{Ang-Bekaert:2002b}
Ang, A. and G.~Bekaert (2002a).
\newblock International asset allocation with regime shifts.
\newblock {\em Review of Financial Studies\/}~{\em 15}, 1137--1187.

\bibitem[\protect\citeauthoryear{Ang and Bekaert}{Ang and
  Bekaert}{2002b}]{Ang-Bekaert:2002a}
Ang, A. and G.~Bekaert (2002b).
\newblock Regime switches in interest rates.
\newblock {\em Journal of Business and Economic Statistics\/}~{\em 20},
  163--182.

\bibitem[\protect\citeauthoryear{Barnard}{Barnard}{1963}]{Barnard:1963}
Barnard, G. (1963).
\newblock Comment on `the spectral analysis of point processes' by m.s.
  bartlett.
\newblock {\em Journal of the Royal Statistical Society (Series B)\/}~{\em 25},
  294.

\bibitem[\protect\citeauthoryear{Beaulieu, Dufour, and Khalaf}{Beaulieu
  et~al.}{2007}]{Beaulieu-Dufour-Khalaf:2007}
Beaulieu, M.-C., J.-M. Dufour, and L.~Khalaf (2007).
\newblock Multivariate tests of mean-variance efficiency with possible
  non-{Gaussian} errors: an exact simulation-based approach.
\newblock {\em Journal of Business and Economic Statistics\/}~{\em 25},
  398--410.

\bibitem[\protect\citeauthoryear{Birnbaum}{Birnbaum}{1974}]{Birnbaum:1974}
Birnbaum, Z. (1974).
\newblock Computers and unconventional test-statistics.
\newblock In F.~Proschan and R.~Serfling (Eds.), {\em Reliability and
  Biometry}, pp.\  441--458. SIAM, Philadelphia.

\bibitem[\protect\citeauthoryear{Carrasco, Hu, and Ploberger}{Carrasco
  et~al.}{2014}]{Carrasco-Hu-Ploberger:2014}
Carrasco, M., L.~Hu, and W.~Ploberger (2014).
\newblock Optimal test for {Markov} switching parameters.
\newblock {\em Econometrica\/}~{\em 82\/}(2), 765--784.

\bibitem[\protect\citeauthoryear{Carter and Steigerwald}{Carter and
  Steigerwald}{2012}]{Carter-Steigerwald:2012}
Carter, A. and D.~Steigerwald (2012).
\newblock Testing for regime switching: a comment.
\newblock {\em Econometrica\/}~{\em 80}, 1809--1812.

\bibitem[\protect\citeauthoryear{Carter and Steigerwald}{Carter and
  Steigerwald}{2013}]{Carter-Steigerwald:2013}
Carter, A. and D.~Steigerwald (2013).
\newblock {Markov} regime-switching tests: asymptotic critical values.
\newblock {\em Journal of Econometric Methods\/}~{\em 2}, 25--34.

\bibitem[\protect\citeauthoryear{Cho and White}{Cho and
  White}{2007}]{Cho-White:2007}
Cho, J. and H.~White (2007).
\newblock Testing for regime switching.
\newblock {\em Econometrica\/}~{\em 75}, 1671--1720.

\bibitem[\protect\citeauthoryear{Davies}{Davies}{1977}]{Davies:1977}
Davies, R. (1977).
\newblock Hypothesis testing when a nuisance parameter is present only under
  the alternative.
\newblock {\em Biometrika\/}~{\em 64}, 274--254.

\bibitem[\protect\citeauthoryear{Davies}{Davies}{1987}]{Davies:1987}
Davies, R. (1987).
\newblock Hypothesis testing when a nuisance parameter is present only under
  the alternative.
\newblock {\em Biometrika\/}~{\em 74}, 33--43.

\bibitem[\protect\citeauthoryear{Davig}{Davig}{2004}]{Davig:2004}
Davig, T. (2004).
\newblock Regime-switching debt and taxation.
\newblock {\em Journal of Monetary Economics\/}~{\em 51}, 837--859.

\bibitem[\protect\citeauthoryear{Dufour}{Dufour}{2006}]{Dufour:2006}
Dufour, J.-M. (2006).
\newblock {Monte Carlo} tests with nuisance parameters: A general approach to
  finite-sample inference and nonstandard asymptotics in econometrics.
\newblock {\em Journal of Econometrics\/}~{\em 133}, 443--477.

\bibitem[\protect\citeauthoryear{Dufour and Khalaf}{Dufour and
  Khalaf}{2001}]{Dufour-Khalaf:2001}
Dufour, J.-M. and L.~Khalaf (2001).
\newblock {Monte Carlo} test methods in econometrics.
\newblock In B.~Baltagi (Ed.), {\em Companion to Theoretical Econometrics}.
  Basil Blackwell, Oxford, UK.

\bibitem[\protect\citeauthoryear{Dufour, Khalaf, Bernard, and Genest}{Dufour
  et~al.}{2004}]{Dufour-Khalaf-Bernard-Genest:2004}
Dufour, J.-M., L.~Khalaf, J.-T. Bernard, and I.~Genest (2004).
\newblock Simulation-based finite-sample tests for heteroskedasticity and arch
  effects.
\newblock {\em Journal of Econometrics\/}~{\em 122}, 317--347.

\bibitem[\protect\citeauthoryear{Dufour, Khalaf, and Voia}{Dufour
  et~al.}{2014}]{Dufour-Khalaf-Voia:2014}
Dufour, J.-M., L.~Khalaf, and M.~Voia (2014).
\newblock Finite-sample resampling-based combined hypothesis tests, with
  applications to serial correlation and predictability.
\newblock {\em Communications in Statistics - Simulation and
  Computation\/}~{\em 44}, 2329--2347.

\bibitem[\protect\citeauthoryear{Dwass}{Dwass}{1957}]{Dwass:1957}
Dwass, M. (1957).
\newblock Modified randomization tests for nonparametric hypotheses.
\newblock {\em Annals of Mathematical Statistics\/}~{\em 28}, 181--187.

\bibitem[\protect\citeauthoryear{Engel and Hamilton}{Engel and
  Hamilton}{1990}]{Engel-Hamilton:1990}
Engel, C. and J.~Hamilton (1990).
\newblock Long swings in the dollar: Are they in the data and do markets know
  it?
\newblock {\em American Economic Review\/}~{\em 80}, 689--713.

\bibitem[\protect\citeauthoryear{Fisher}{Fisher}{1932}]{Fisher:1932}
Fisher, R. (1932).
\newblock {\em Statistical Methods for Research Workers}.
\newblock Oliver and Boyd, Edinburgh.

\bibitem[\protect\citeauthoryear{Garcia}{Garcia}{1998}]{Garcia:1998}
Garcia, R. (1998).
\newblock Asymptotic null distribution of the likelihood ratio test in {Markov}
  switching models.
\newblock {\em International Economic Review\/}~{\em 39}, 763--788.

\bibitem[\protect\citeauthoryear{Garcia and Perron}{Garcia and
  Perron}{1996}]{Garcia-Perron:1996}
Garcia, R. and P.~Perron (1996).
\newblock An analysis of the real interest rate under regime shifts.
\newblock {\em Review of Economics and Statistics\/}~{\em 78}, 111--125.

\bibitem[\protect\citeauthoryear{Gouri\'{e}roux and Monfort}{Gouri\'{e}roux and
  Monfort}{1995}]{Gourieroux-Monfort:1995}
Gouri\'{e}roux, C. and A.~Monfort (1995).
\newblock {\em Statistics and Econometric Models}, Volume~1.
\newblock Cambridge University Press.

\bibitem[\protect\citeauthoryear{Guidolin}{Guidolin}{2011}]{Guidolin:2011}
Guidolin, M. (2011).
\newblock Markov switching models in empirical finance.
\newblock In D.~Drukker (Ed.), {\em Missing Data Methods: Time-Series Methods
  and Applications (Advances in Econometrics, Volume 27 Part 2)}. Emerald Group
  Publishing Limited.

\bibitem[\protect\citeauthoryear{Hamilton}{Hamilton}{1989}]{Hamilton:1989}
Hamilton, J. (1989).
\newblock A new approach to the economic analysis of nonstationary time series
  and the business cycle.
\newblock {\em Econometrica\/}~{\em 57}, 357--384.

\bibitem[\protect\citeauthoryear{Hamilton}{Hamilton}{1994}]{Hamilton:1994}
Hamilton, J. (1994).
\newblock {\em Time Series Analysis}.
\newblock Princeton University Press, Princeton, New Jersey.

\bibitem[\protect\citeauthoryear{Hamilton}{Hamilton}{2016}]{Hamilton:2016}
Hamilton, J. (2016).
\newblock Macroeconomic regimes and regime shifts.
\newblock In J.~Taylor and H.~Uhlig (Eds.), {\em Handbook of Macroeconomics,
  Vol. 2}. Elsevier Science Publishers B.V.

\bibitem[\protect\citeauthoryear{Hamilton and Susmel}{Hamilton and
  Susmel}{1994}]{Hamilton-Susmel:1994}
Hamilton, J. and R.~Susmel (1994).
\newblock Autoregressive conditional heteroskedasticity and changes in regime.
\newblock {\em Journal of Econometrics\/}~{\em 64}, 307--333.

\bibitem[\protect\citeauthoryear{Hansen}{Hansen}{1992}]{Hansen:1992}
Hansen, B. (1992).
\newblock The likelihood ratio test under nonstandard conditions: Testing the
  {Markov} switching model of {GNP}.
\newblock {\em Journal of Applied Econometrics\/}~{\em 7}, S61--S82.

\bibitem[\protect\citeauthoryear{Hansen}{Hansen}{1996}]{Hansen:1996}
Hansen, B. (1996).
\newblock Erratum: The likelihood ratio test under nonstandard conditions:
  Testing the {Markov} switching model of {GNP}.
\newblock {\em Journal of Applied Econometrics\/}~{\em 11}, 195--198.

\bibitem[\protect\citeauthoryear{Hodges and Lehmann}{Hodges and
  Lehmann}{1963}]{Hodges-Lehmann:1963}
Hodges, J. and E.~Lehmann (1963).
\newblock Estimates of location based on rank tests.
\newblock {\em The Annals of Mathematical Statistics\/}~{\em 35}, 598--611.

\bibitem[\protect\citeauthoryear{Kim and Nelson}{Kim and
  Nelson}{1999}]{Kim-Nelson:1999}
Kim, C. and C.~Nelson (1999).
\newblock Has the {U.S.} economy become more stable? a {Bayesian} approach
  based on a {Markov-switching} model of the business cycle.
\newblock {\em Review of Economic and Statistics\/}~{\em 81}, 608--616.

\bibitem[\protect\citeauthoryear{Lee and Chesher}{Lee and
  Chesher}{1986}]{Lee-Chesher:1986}
Lee, L.-F. and A.~Chesher (1986).
\newblock Specification testing when score statistics are identically zero.
\newblock {\em Journal of Econometrics\/}~{\em 31}, 121--149.

\bibitem[\protect\citeauthoryear{MacKinnon}{MacKinnon}{2009}]{MacKinnon:2009}
MacKinnon, J. (2009).
\newblock Bootstrap hypothesis testing.
\newblock In D.~Belsley and J.~Kontoghiorghes (Eds.), {\em Handbook of
  Computational Econometrics}, pp.\  183--213. Wiley.

\bibitem[\protect\citeauthoryear{McConnell and Perez-Quiros}{McConnell and
  Perez-Quiros}{2000}]{McConnell-Perez-Quiros:2000}
McConnell, M. and G.~Perez-Quiros (2000).
\newblock Output fluctuations in the {United States}: What has changed since
  the early 1980's?
\newblock {\em American Economic Review\/}~{\em 90}, 1464--1476.

\bibitem[\protect\citeauthoryear{Pearson}{Pearson}{1933}]{Pearson:1933}
Pearson, K. (1933).
\newblock On a method of determining whether a sample of size $n$ supposed to
  have been drawn from a parent population having a known probability integral
  has probably been drawn at random.
\newblock {\em Biometrika\/}~{\em 25}, 379--410.

\bibitem[\protect\citeauthoryear{Psaradakis and Sola}{Psaradakis and
  Sola}{1998}]{Psaradakis-Sola:1998}
Psaradakis, Z. and M.~Sola (1998).
\newblock Finite-sample properties of the maximum likelihood estimator in
  autoregressive models with {Markov} switching.
\newblock {\em Journal of Econometrics\/}~{\em 86}, 369--386.

\bibitem[\protect\citeauthoryear{Rossi}{Rossi}{2014}]{Rossi:2014}
Rossi, P. (2014).
\newblock {\em Bayesian Non- and Semi-parametric Methods and Applications}.
\newblock Princeton University Press.

\bibitem[\protect\citeauthoryear{Savin}{Savin}{1984}]{Savin:1984}
Savin, N. (1984).
\newblock Multiple hypothesis testing.
\newblock In Z.~Griliches and M.~Intriligator (Eds.), {\em Handbook of
  Econometrics}, pp.\  827--879. North-Holland, Amsterdam.

\bibitem[\protect\citeauthoryear{Timmermann}{Timmermann}{2000}]{Timmermann:2000}
Timmermann, A. (2000).
\newblock Moments of {Markov} switching models.
\newblock {\em Journal of Econometrics\/}~{\em 96}, 75--111.

\bibitem[\protect\citeauthoryear{Timmermann}{Timmermann}{2001}]{Timmermann:2001}
Timmermann, A. (2001).
\newblock Structural breaks, incomplete information and stock prices.
\newblock {\em Journal of Business and Economic Statistics\/}~{\em 19},
  299--315.

\bibitem[\protect\citeauthoryear{Tippett}{Tippett}{1931}]{Tippett:1931}
Tippett, L. (1931).
\newblock {\em The Method of Statistics}.
\newblock Williams \& Norgate, London.

\bibitem[\protect\citeauthoryear{Watson and Engle}{Watson and
  Engle}{1985}]{Watson-Engle:1985}
Watson, M. and R.~Engle (1985).
\newblock Testing for regression coefficient stability with a stationary
  {AR(1)} alternative.
\newblock {\em Review of Economics and Statistics\/}~{\em 67}, 341--346.

\bibitem[\protect\citeauthoryear{Wilkinson}{Wilkinson}{1951}]{Wilkinson:1951}
Wilkinson, B. (1951).
\newblock A statistical consideration in psychological research.
\newblock {\em Psychology Bulletin\/}~{\em 48}, 156--158.

\end{thebibliography}

{\renewcommand{\baselinestretch}{1} }

\begin{table}[p]
\begin{center}
Table 1. Coefficients of approximate distribution functions \medskip
\par
\begin{tabular}{lcrrcrrcrrcrr}
\hline\hline
&  & \multicolumn{2}{c}{$\hat{F}_M$} &  & \multicolumn{2}{c}{$\hat{F}_V$} & 
& \multicolumn{2}{c}{$\hat{F}_S$} &  & \multicolumn{2}{c}{$\hat{F}_K$} \\ 
\cline{3-4}\cline{6-7}\cline{9-10}\cline{12-13}
&  & $\hat{\gamma}_0$ & $\hat{\gamma}_0$ &  & $\hat{\gamma}_0$ & $\hat{\gamma%
}_0$ &  & $\hat{\gamma}_0$ & $\hat{\gamma}_0$ &  & $\hat{\gamma}_0$ & $\hat{%
\gamma}_0$ \\ \hline
&  &  &  &  &  &  &  &  &  &  &  &  \\[-2.0ex] 
$T$=50 &  & -16.178 & 8.380 &  & -7.700 & 0.879 &  & -1.944 & 8.423 &  & 
-2.191 & 5.106 \\ 
$T$=100 &  & -23.041 & 12.125 &  & -10.923 & 1.253 &  & -1.975 & 11.614 &  & 
-2.101 & 6.538 \\ 
$T$=150 &  & -28.289 & 14.961 &  & -13.394 & 1.539 &  & -1.995 & 14.128 &  & 
-2.068 & 7.690 \\ 
$T$=200 &  & -32.719 & 17.348 &  & -15.484 & 1.781 &  & -2.012 & 16.311 &  & 
-2.051 & 8.680 \\ 
$T$=250 &  & -36.653 & 19.463 &  & -17.312 & 1.992 &  & -2.021 & 18.197 &  & 
-2.046 & 9.597 \\ \hline\hline
\end{tabular}%
\end{center}
\par
Note: The entries are the coefficients of the approximate distribution
functions in (\ref{logi}) used to compute the first-level p-values in the
test combination procedure. The coefficients are obtained by NLS with one
million simulated samples for each sample size, $T$.
\end{table}

\begin{table}[p]
\begin{center}
Table 2. Empirical size of tests for Markov-switching \medskip
\par
\begin{tabular}{lrrrrrrr}
\hline\hline
&  & \multicolumn{2}{c}{$\phi=0.1$} &  & \multicolumn{2}{c}{$\phi=0.9$} & 
\\ \cline{3-4}\cline{6-7}
Test &  & $T=100$ & $T=200$ &  & $T=100$ & $T=200$ &  \\ \hline
LMC$_{\min}$ &  & 5.3 & 4.6 &  & 4.9 & 4.4 &  \\ 
LMC$_{\times}$ &  & 5.2 & 4.9 &  & 4.7 & 4.4 &  \\ 
MMC$_{\min}$ &  & 0.6 & 0.6 &  & 0.8 & 1.0 &  \\ 
MMC$_{\times}$ &  & 0.2 & 0.5 &  & 0.9 & 1.2 &  \\ 
supTS &  & 4.8 & 5.1 &  & 6.0 & 4.5 &  \\ 
expTS &  & 6.8 & 6.2 &  & 5.4 & 6.9 &  \\ \hline\hline
\end{tabular}%
\end{center}
\par
Note: The DGP is an AR(1) model and the nominal level is 5\%. LMC and MMC
stand for the local and maximized MC procedures, respectively. The subscript
``$\min$" means that the first-level p-values are combined via their
minimum, while the subscript ``$\times$" means that they are combined via
their product. The supTS and expTS tests refer to the supremum-type and
exponential-type tests of \citet{Carrasco-Hu-Ploberger:2014}.
\end{table}

\begin{table}[p]
\begin{center}
Table 3. Empirical power of tests for Markov-switching with $\phi=0.1$
\medskip
\par
\begin{tabular}{lrrrrrrrrr}
\hline\hline
&  & \multicolumn{2}{c}{$(p_{11}, p_{22})=(0.9, 0.9)$} &  & 
\multicolumn{2}{c}{$(p_{11}, p_{22})=(0.9, 0.5)$} &  & \multicolumn{2}{c}{$%
(p_{11}, p_{22})=(0.9, 0.1)$} \\ \cline{3-4}\cline{6-7}\cline{9-10}
Test &  & $T=100$ & $T=200$ &  & $T=100$ & $T=200$ &  & $T=100$ & $T=200$ \\ 
\hline
\multicolumn{10}{l}{$\Delta \mu =2, \Delta \sigma=0$} \\ 
LMC$_{\min}$ &  & 5.8 & 4.7 &  & 14.4 & 26.7 &  & 20.1 & 39.2 \\ 
LMC$_{\times}$ &  & 6.8 & 4.6 &  & 12.5 & 23.4 &  & 19.0 & 36.6 \\ 
MMC$_{\min}$ &  & 0.4 & 0.3 &  & 1.9 & 7.6 &  & 2.8 & 15.5 \\ 
MMC$_{\times}$ &  & 0.6 & 0.3 &  & 2.3 & 7.1 &  & 3.1 & 13.9 \\ 
supTS &  & 24.3 & 49.9 &  & 23.8 & 47.0 &  & 24.4 & 45.6 \\ 
expTS &  & 15.6 & 25.4 &  & 24.6 & 47.1 &  & 28.9 & 52.3 \\ 
\multicolumn{10}{l}{$\Delta \mu =0, \Delta \sigma=1$} \\ 
LMC$_{\min}$ &  & 39.4 & 62.0 &  & 48.4 & 72.6 &  & 40.0 & 55.7 \\ 
LMC$_{\times}$ &  & 42.6 & 64.3 &  & 49.4 & 73.2 &  & 41.3 & 55.5 \\ 
MMC$_{\min}$ &  & 15.5 & 39.0 &  & 28.1 & 55.2 &  & 21.2 & 40.7 \\ 
MMC$_{\times}$ &  & 17.1 & 43.2 &  & 27.3 & 52.8 &  & 19.9 & 39.8 \\ 
supTS &  & 32.4 & 58.0 &  & 29.9 & 46.4 &  & 22.8 & 30.4 \\ 
expTS &  & 40.1 & 62.6 &  & 43.9 & 68.3 &  & 34.4 & 52.4 \\ 
\multicolumn{10}{l}{$\Delta \mu =2, \Delta \sigma=1$} \\ 
LMC$_{\min}$ &  & 52.3 & 84.0 &  & 82.1 & 98.8 &  & 78.5 & 96.3 \\ 
LMC$_{\times}$ &  & 46.6 & 75.4 &  & 82.8 & 98.9 &  & 80.0 & 96.3 \\ 
MMC$_{\min}$ &  & 21.7 & 51.9 &  & 57.0 & 92.5 &  & 57.1 & 89.5 \\ 
MMC$_{\times}$ &  & 23.0 & 49.0 &  & 61.3 & 93.5 &  & 59.6 & 90.2 \\ 
supTS &  & 72.7 & 96.2 &  & 80.8 & 96.9 &  & 65.5 & 89.7 \\ 
expTS &  & 75.6 & 97.0 &  & 86.6 & 99.4 &  & 78.2 & 96.2 \\ \hline\hline
\end{tabular}%
\end{center}
\par
Note: The DGP is model (\ref{ar2}) with $\phi=0.1$ and the nominal level is
5\%. LMC and MMC stand for the local and maximized MC procedures,
respectively. The subscript ``$\min$" means that the first-level p-values
are combined via their minimum, while the subscript ``$\times$" means that
they are combined via their product. The supTS and expTS tests refer to the
supremum-type and exponential-type tests of %
\citet{Carrasco-Hu-Ploberger:2014}.
\end{table}

\begin{table}[p]
\begin{center}
Table 4. Empirical power of tests for Markov-switching with $\phi=0.9$
\medskip
\par
\begin{tabular}{lrrrrrrrrr}
\hline\hline
&  & \multicolumn{2}{c}{$(p_{11}, p_{22})=(0.9, 0.9)$} &  & 
\multicolumn{2}{c}{$(p_{11}, p_{22})=(0.9, 0.5)$} &  & \multicolumn{2}{c}{$%
(p_{11}, p_{22})=(0.9, 0.1)$} \\ \cline{3-4}\cline{6-7}\cline{9-10}
Test &  & $T=100$ & $T=200$ &  & $T=100$ & $T=200$ &  & $T=100$ & $T=200$ \\ 
\hline
\multicolumn{10}{l}{$\Delta \mu =2, \Delta \sigma=0$} \\ 
LMC$_{\min}$ &  & 15.5 & 21.8 &  & 14.5 & 22.2 &  & 14.8 & 24.5 \\ 
LMC$_{\times}$ &  & 15.2 & 23.0 &  & 14.4 & 20.9 &  & 14.9 & 25.9 \\ 
MMC$_{\min}$ &  & 3.8 & 7.9 &  & 3.7 & 6.9 &  & 3.3 & 7.9 \\ 
MMC$_{\times}$ &  & 3.6 & 7.4 &  & 3.8 & 9.1 &  & 3.2 & 9.7 \\ 
supTS &  & 8.4 & 12.5 &  & 11.9 & 18.2 &  & 20.7 & 45.6 \\ 
expTS &  & 21.7 & 32.6 &  & 22.1 & 33.5 &  & 25.6 & 43.2 \\ 
\multicolumn{10}{l}{$\Delta \mu =0, \Delta \sigma=1$} \\ 
LMC$_{\min}$ &  & 37.8 & 64.7 &  & 48.1 & 70.9 &  & 38.9 & 61.7 \\ 
LMC$_{\times}$ &  & 40.9 & 68.1 &  & 48.5 & 72.8 &  & 40.1 & 62.7 \\ 
MMC$_{\min}$ &  & 17.1 & 42.2 &  & 27.8 & 55.5 &  & 22.6 & 47.3 \\ 
MMC$_{\times}$ &  & 19.9 & 43.8 &  & 28.1 & 55.4 &  & 22.2 & 45.2 \\ 
supTS &  & 32.2 & 67.4 &  & 30.0 & 50.3 &  & 20.0 & 34.1 \\ 
expTS &  & 54.1 & 84.7 &  & 52.8 & 78.6 &  & 41.9 & 65.3 \\ 
\multicolumn{10}{l}{$\Delta \mu =2, \Delta \sigma=1$} \\ 
LMC$_{\min}$ &  & 40.9 & 64.4 &  & 65.7 & 88.8 &  & 70.9 & 89.0 \\ 
LMC$_{\times}$ &  & 42.1 & 65.8 &  & 67.6 & 91.2 &  & 72.0 & 90.6 \\ 
MMC$_{\min}$ &  & 16.8 & 37.5 &  & 41.8 & 76.6 &  & 50.2 & 77.3 \\ 
MMC$_{\times}$ &  & 19.3 & 44.1 &  & 46.4 & 83.2 &  & 53.3 & 82.1 \\ 
supTS &  & 34.6 & 62.9 &  & 53.2 & 79.8 &  & 58.6 & 82.3 \\ 
expTS &  & 53.9 & 77.9 &  & 75.1 & 94.7 &  & 77.4 & 94.2 \\ \hline\hline
\end{tabular}%
\end{center}
\par
Note: The DGP is model (\ref{ar2}) with $\phi=0.9$ and the nominal level is
5\%. LMC and MMC stand for the local and maximized MC procedures,
respectively. The subscript ``$\min$" means that the first-level p-values
are combined via their minimum, while the subscript ``$\times$" means that
they are combined via their product. The supTS and expTS tests refer to the
supremum-type and exponential-type tests of %
\citet{Carrasco-Hu-Ploberger:2014}.
\end{table}

\begin{table}[p]
\begin{center}
Table 5. MC test results: U.S. real GNP growth \medskip
\par
\begin{tabular}{lrccrrrrcr}
\hline\hline
Test &  & p-value &  & $\phi_1$ & $\phi_2$ & $\phi_3$ & $\phi_4$ &  & $|z|$
\\ \hline
\multicolumn{10}{l}{1952Q2 -- 1984Q4} \\ 
LMC$_{\min}$ &  & 0.57 &  & 0.31 & 0.13 & -0.12 & -0.09 &  & 1.50 \\ 
LMC$_{\times}$ &  & 0.57 &  & 0.31 & 0.13 & -0.12 & -0.09 &  & 1.50 \\ 
MMC$_{\min}$ &  & 1.00 &  & 0.48 & 0.20 & -0.23 & -0.16 &  & 1.23 \\ 
MMC$_{\times}$ &  & 1.00 &  & 0.38 & 0.30 & -0.28 & -0.09 &  & 1.32 \\ 
\multicolumn{10}{l}{1952Q2 -- 2010Q4} \\ 
LMC$_{\min}$ &  & 0.01 &  & 0.34 & 0.12 & -0.08 & -0.07 &  & 1.59 \\ 
LMC$_{\times}$ &  & 0.01 &  & 0.34 & 0.12 & -0.08 & -0.07 &  & 1.59 \\ 
MMC$_{\min}$ &  & 0.05 &  & 0.43 & 0.09 & 0.05 & 0.05 &  & 1.33 \\ 
MMC$_{\times}$ &  & 0.06 &  & 0.46 & 0.08 & 0.05 & 0.02 &  & 1.41 \\ 
\hline\hline
&  &  &  &  &  &  &  &  &  \\[-2.0ex] 
&  &  &  &  &  &  &  &  & 
\end{tabular}%
\end{center}
\par
Note: LMC and MMC stand for the local and maximized MC procedures,
respectively. The subscript ``$\min$" means that the first-level p-values
are combined via their minimum, while the subscript ``$\times$" means that
they are combined via their product. Entries under $|z|$ are the smallest
moduli of the roots of the autoregressive polynomial for the corresponding
line.
\end{table}

\end{document}